\documentclass{aa}  

\usepackage{txfonts}
\usepackage{url}
\usepackage{hyperref}
\usepackage{orcidlink}
\usepackage{graphicx}
\usepackage{txfonts}
\newcommand{\dg}{^{\circ}}
\newcommand{\multfact}{0.63}
\newcommand{\Nov}{Paper I}

\begin{document}

   \title{Repeated pattern of $\gamma$-ray flares in the blazar \object{PKS 1502+106} coincident with the IC190730A neutrino event}
   \titlerunning{Repeated $\gamma$-ray flares in PKS 1502+106 coincident with IC190730A}

   \subtitle{Growing evidence of structured jets as neutrino sources}

   \author{Dmitry Blinov
          \inst{1,2}\orcidlink{0000-0003-0611-5784} 
          \and
          Polina Novikova\inst{3}\orcidlink{0009-0002-9081-5563} 
          }

   \institute{Institute of Astrophysics, Foundation for Research and Technology-Hellas, GR-70013 Heraklion, Greece
   \and
   Department of Physics, University of Crete, GR-70013, Heraklion, Greece\\
              \email{blinov@ia.forth.gr}
         \and
             St. Petersburg University, 7/9 Universitetskaya nab., St.Petersburg, 199034, Russia\\
             }

   \date{Received December 10, 2024; accepted YY}

 
  \abstract
   {It has been demonstrated that at least 10 percent of the brightest blazars in the fourth Fermi-LAT catalog of $\gamma$-ray sources exhibit repeating patterns of $\gamma$-ray flares. These events may be associated with the presence of a non-uniform sheath surrounding a fast jet spine in some blazars. Theoretical models suggest that such a sheath could facilitate neutrino production in these structured jets.}
   {We aim to test the marginal statistical evidence previously reported for a connection between repeating patterns of $\gamma$-ray flares in blazars and high-energy neutrino events that are positionally consistent with these sources.}
   {We identified a repeating pattern of flares in the $\gamma$-ray light curve of the blazar PKS 1502+106, which lies within the 50\% uncertainty region of the IC190730A neutrino candidate event. This occurrence is combined with two other high-energy ($\ge 200$ TeV) neutrino events from ICECAT-1, which  arrived in both positional and temporal coincidence with two blazars exhibiting ongoing repeating flare patterns. We conducted a Monte Carlo simulation  to evaluate the likelihood of accidental coincidences between the repeating flare patterns and neutrino events, accounting for potential unrecognized systematic uncertainties in the arrival directions of the ICECAT-1 events.}
   {Our findings indicate the probability of a random coincidence, in both time and arrival direction for three high-energy neutrino candidates and three blazars with ongoing recurring patterns of $\gamma$-ray flares, is $1.56\times 10^{-3}$ ($3.2\sigma$).}
   {}


   \maketitle
%

\section{Introduction} \label{introduction}

The IC170922A event, a 290 TeV neutrino that is likely astrophysical in origin, was shown to be directionally consistent with blazar TXS 0506+056. A multi-wavelength campaign revealed increased emission across the electromagnetic spectrum, suggesting an astrophysical link, with a $3\sigma$ significance \citep{Aartsen2018}. This discovery has fueled discussions about blazars as potential sources of astrophysical neutrinos. Multiple studies have identified statistical associations between neutrino events and blazars \citep[e.g.,][]{Plavin2020,Plavin2023,Buson2022}, although such connections remain debated, even when the  adopted datasets are the same \citep{Bellenghi2023,Abbasi2023a}. The evidence remains inconclusive due to the poor localization of neutrinos and the relatively weak signal compared to background noise in the data. While blazars may account for only a small fraction of astrophysical neutrinos \citep{Smith2021}, it is also evident that only a minority of active galactic nuclei (AGNs) produce neutrinos \citep{Abbasi2023a}. This supports the hypothesis that a specific subclass of blazars might be responsible for significant neutrino production \citep[e.g.,][]{Giommi2021,Padovani2022}.

We investigated the potential connection between neutrino events and recurring patterns of $\gamma$-ray flares in blazars, as discussed in \cite{Novikova2023} (hereafter, \Nov). These patterns have non-trivial shapes consisting of multiple flares and reappear in the light curve. They are likely produced when emission features propagate through the fast spine of the jet, surrounded by a slower sheath, as demonstrated in \cite{Blinov2021}. Such structured jets not only enhance inverse Compton $\gamma$-ray emission \citep{Ghisellini2005} but they are also expected to generate a higher neutrino luminosity, compared to jets without a sheath \citep{Tavecchio2014}. The increased neutrino output in the spine arises due to the photo-meson interaction of high-energy protons with the intensified soft target photon field provided by the sheath \citep{Tavecchio2015}.

In a previous work, we  found that at least 10\% of $\gamma$-ray bright blazars produce repeated patterns of flares (\Nov), indicating the presence of a spine-sheath structure in their jets. Among these ten blazars displaying repeated patterns (referred to as "repeaters"), five appear to be positionally consistent with neutrino events that occurred during an ongoing repetition of the pattern in the respective source. However, the sample of five neutrino occurrences includes different types of events (track-like, cascade-like, and a low-energy neutrino flare) detected by two telescopes (IceCube and Baikal-GVD). This heterogeneity complicates the assessment of the statistical significance of the connection between $\gamma$-ray patterns and neutrinos based on this sample. To address this limitation, in \Nov,{} we focused on two track-like neutrino events, IC170922A and IC150926A, from the homogeneous \citep[ICECAT-1,][]{Abbasi2023b} catalog. These events originated from directions consistent with the blazars J0509.4+0542 and J1256.1$-$0547, respectively,  detected during ongoing repeated patterns of $\gamma$-ray flares. Based on these two events, we obtained a $2.8\sigma$ significance for the association between neutrinos and repeated patterns.

In this paper, we report the discovery of one additional repeated pattern in the blazar PKS 1502+106 (a.k.a. 4FGL J1504.4+1029). During one occurrence of this pattern, a neutrino event that is positionally consistent with the source was detected.  PKS 1502+106 ($z = 1.838$) is one of the brightest sources in the Fermi Large Area Telescope fourth source catalog \citep[4FGL,][]{Abdollahi2020}. It has been extensively studied across the electromagnetic spectrum \citep[e.g.,][]{Karamanavis2016a,Karamanavis2016b} and previously discussed as a potential neutrino source \citep[e.g.,][]{Britzen2021,Oikonomou2021}. Here, we report the discovered $\gamma$-ray flare pattern in this source and estimate its significance. We incorporate this new event into the analysis of \Nov{} and re-evaluate the significance of the connection between repeated patterns and neutrinos.

\section{Data and data reduction} \label{data}

The Fermi Large Area Telescope \citep[LAT,][]{Atwood2009} data were processed in the 100 MeV to 300 GeV range using the unbinned likelihood analysis from Fermitools (v. 1.2.23) via Conda. We applied the instrument response function P8R3\_SOURCE\_V3, selecting source class photons (evclass=128, evtype=3) within a $15\dg$ region of interest (ROI) centered on a blazar. To exclude the Earth limb background, we limited the zenith angle to $<90\dg$. The Galactic interstellar emission was modeled with gll\_iem\_v07, while extragalactic and residual backgrounds used the isotropic template iso\_P8R3\_SOURCE\_V3\_v1.txt. All sources within $15\dg$ of the blazar from the 4FGL were included, with photon fluxes fixed for sources beyond $10\dg$. Spectral shapes for all ROI targets were fixed. A detection threshold of $TS = 10$ ($\sim3\sigma$) was applied. Systematic uncertainties in the LAT area were under 10\% \citep{Ackermann2012} and only statistical errors were considered, due to the dominance over short timescales.

The IceCube neutrino telescope \citep{IceCube2013} detects high-energy neutrino events in two forms: cascades and tracks. Track-like events are particularly valuable for associating neutrinos with astrophysical sources due to their smaller directional uncertainties, typically around $1^\circ$. The likelihood of a neutrino originating from an astrophysical source increases with energy, and neutrinos above 200 TeV have a $>50\%$ chance of being astrophysical. Following the selection criteria from \cite{Plavin2020} and \cite{Hovatta2021}, we selected neutrinos with the energy $E \geq 200$ TeV from the IceCube Event Catalog of Alert Tracks \citep[ICECAT-1,][]{Abbasi2023b}. We applied a positional accuracy cut, limiting the 90\% containment area to $\Omega_{90} \leq 10 \, \text{deg}^2$. This resulted in 54 neutrinos detected between September 2, 2011 and December 9, 2020.

\section{Repeated pattern in PKS 1502+106} \label{sec:patt}

\begin{figure*}
     \includegraphics[width=0.99\hsize]{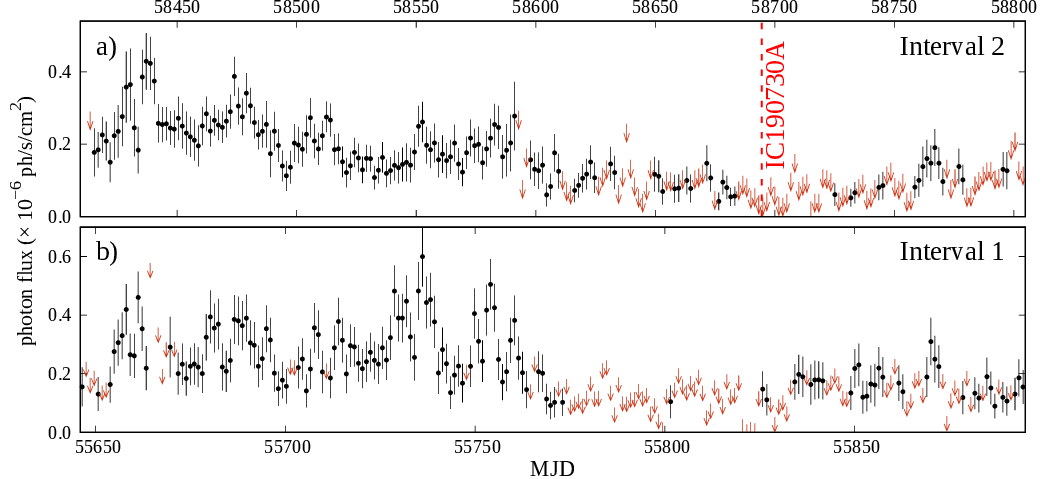}
        \caption{Two intervals of the $\gamma$-ray light curve of PKS 1502+106 exhibiting the repeated pattern of flares for: a) Interval between 18 October 2018 and 17 November 2019, processed with a 3.35 d integration; b) Interval between 26 March 2011 and 30 November 2011, processed with a 2.11 d integration. Both curves are over-sampled in such a way that the time difference between two consecutive points equals half of the integration time. The dashed red line marks the arrival time of IC190730A.}
        \label{fig:pattern}
  \end{figure*}

In our previous study (\Nov), we performed an automated search for repeated patterns of flares in 14-year-long $\gamma$-ray light curves of the 100 brightest blazars in the 4FGL catalog. However, during subsequent visual inspection, we discovered that some patterns that did not conform to the algorithm's initial assumptions were overlooked. To address this limitation, we identified blazars within the uncertainty regions of the 54 selected ICECAT-1 neutrinos that are among the 100 brightest $\gamma$-ray blazars in the 4FGL catalog. We then obtained the Fermi-LAT light curves for these blazars (as described in Sect.~\ref{data}) and conducted a visual inspection to search for patterns in these curves. Using this approach, we identified a repeated episode of activity in PKS 1502+106, shown in Fig.\ref{fig:pattern}.
The timescale of interval 2 is compressed by a factor of 0.63 compared to the timescale of interval 1. This factor corresponds to the ratio of Doppler factors of the moving features responsible for the patterns \citep{Blinov2021}. The transformation was determined by minimizing the Time-Warp Edit Distance \citep{Lin2012} between the two repetitions. In Appendix~\ref{app:A}, we demonstrate that these two intervals of the light curve provide the best match over the period covered by ICECAT-1.

\section{Association of neutrinos with patterns} \label{neutrino}

During the second repetition of the identified pattern, the track-like neutrino event IC190730A was detected by IceCube. The moment of its arrival is indicated by the red dashed line in Fig.~\ref{fig:pattern}. Since this event is included in ICECAT-1 and meets the selection criteria used in the analysis of \Nov{}, we incorporated it into the sample and re-evaluated the statistical significance of the association between neutrino events and repeated patterns.

To assess this significance, we focused on the uniformly selected sample of 54 track-like IceCube neutrino events (Sect.~\ref{data}) and applied the method outlined by \cite{Plavin2020} and \cite{Hovatta2021}. For each neutrino in the sample, we transformed the 90$\%$ coordinate-wise statistical uncertainties $\Delta^{\rm stat}_i$ in right ascension (RA) and declination (DEC), as provided in ICECAT-1, to derive two-dimensional (2D)\ 90$\%$ coverage regions. This was achieved by multiplying the coordinate-wise errors by the ratio of 90$\%$ quantiles of 2D and 1D Gaussian distributions: $\frac{\sqrt{-{\rm log}(1-0.9)}}{{\rm erf}^{-1}(0.9)} \approx 1.3$ \citep{Plavin2020}.

In addition to statistical uncertainties, a systematic positional uncertainty, $\Delta \Psi$, must be considered. This uncertainty primarily arises from limitations in the understanding of the optical properties of the ice \citep{IceCube2013}. Since $\Delta \Psi$ values are not available for all neutrino events, we estimated this parameter using our data. Following standard practices in particle and astroparticle physics \citep[e.g.,][]{Abbasi2012}, we optimized the signal by tuning $\Delta \Psi$. This optimization was conducted across a range of $\Delta \Psi$ values, from $0.1\dg$ to $1.0\dg$, in increments of $0.01\dg$. The total positional uncertainty for each neutrino, incorporating both statistical and systematic components, was calculated as: $\sqrt{(\Delta^{\rm stat}_i)^2 + (\Delta \Psi)^2}$.

We defined a statistic, $\nu$, as the number of blazars falling within these uncertainty regions and exhibiting ongoing repeated patterns during the corresponding neutrino arrival times. For each $\Delta \Psi$, we calculated the observed statistic $\nu_{\rm obs}$ based on the actual coordinates of the neutrino events. In this study, $\nu_{\rm obs}$ is equal to 3, corresponding to  events IC170922A, IC150926A, and IC190730A, which are associated with TXS 0506+056, 3C 279, and PKS 1502+106, respectively. To estimate the probability of these associations occurring by chance, we performed a Monte Carlo (MC) simulation. In the simulation, we randomized the RA of the neutrinos while keeping the DEC and arrival times unchanged, as IceCube's sensitivity primarily depends on the zenith angle \citep{Aartsen2017b}. Additionally, we randomized the intervals of repeated patterns by shifting them to random times within the ICECAT-1 observation window, preserving their original durations. We conducted $N=10^7$ MC trials and calculated the statistic $\nu_i$ for each trial. The number of trials, $M$, where $\nu_i \ge \nu_{\rm obs}$ was determined, and the probability of $\nu_{\rm obs}$ being a random occurrence was calculated as \citep{Davison1997}:
\begin{equation} \label{eq:1}
{\rm p-value} = \frac{M+1}{N+1}.
\end{equation}
By varying $\Delta \Psi$ and minimizing the p-value, we obtained the pre-trial p-value. To correct for multiple comparisons, we performed an additional layer of MC simulations using the generated neutrino samples and their corresponding $\nu_i$ values. Each random sample was treated as an observation and compared with $N=10^7-1$ other random samples. The pre-trial p-value for each random sample was minimized across $\Delta \Psi$, and the number of simulations, $M$, was determined for cases where the pre-trial p-value was less than or equal to the one obtained in the real data. Finally, using equation~\ref{eq:1}, we calculated the post-trial p-value = $1.57 \times 10^{-3}$  ($3.2\sigma$), unaffected by multiple comparisons. This value provides the revised significance estimate for the association between neutrinos with repeated patterns.

\section{Discussion and conclusions}

The significance estimate for the three events analyzed here has been increased to $3.2\sigma$, compared to the $2.8\sigma$ significance reported for two events in \Nov. Notably, these significance levels can also be approximated using a simpler method. If the repeated patterns were randomly distributed over time, relative to the neutrino arrival times for a given blazar, the probability of coincidence for two events would roughly correspond the ratio of the total duration of the patterns within the ICECAT-1 time range to the total duration of the time range itself. For the three blazars, TXS 0506+056, 3C 279, and PKS 1502+106, these ratios are 0.05, 0.11, and 0.19, respectively. The probability of neutrino events coinciding with ongoing patterns in the first two or all three of these blazars can be calculated by multiplying the respective ratios. This yields p-values of $5.4 \times 10^{-3}$ and $10^{-3}$, corresponding to significance levels of $2.78\sigma$ and $3.29\sigma$, respectively, which are in good agreement with the MC-based estimates. Using this approach and the average ratio of the total pattern duration to the ICECAT-1 time range duration $0.12$, we have estimated that achieving a $5\sigma$ association between repeated patterns and track-like neutrinos from ICECAT-1 would require at least seven neutrino events arriving in coincidence with repeated patterns ($\text{p-value} = 0.12^7 = 3.6 \times 10^{-7}$).

It appears unlikely that the existing sample of repeated patterns can be more than doubled, as most of the brightest $\gamma$-ray blazars have already been analyzed, and searches in fainter sources are limited by the sensitivity of Fermi-LAT. However, it is important to note that this analysis only considers track-like neutrino events from ICECAT-1. When including three additional neutrino events associated with repeated patterns (see Sect. \ref{introduction} and \Nov), there are a total of six events coinciding with repeated patterns in 6 out of 11 blazars known to exhibit such patterns. This broader set of events could potentially yield a higher significance than what is estimated in this work. Additionally, analyzing low-energy neutrino flux excesses in repeaters could be advantageous, as our pattern intervals predict when such excesses are expected to occur.

An alternative approach for testing the possible connection between spine-sheath jets and neutrino events involves identifying such jets using other proxies. For instance, the jet in TXS 0506+056 has been shown to exhibit limb brightening, likely associated with a spine-sheath structure, which may explain the neutrino flux of this blazar \citep{Ros2020}. Implementing a dedicated, high-sensitivity very long baseline interferometry imaging program could prove valuable to systematic identifications of structured jets.

The increasing number of identified neutrino events coinciding with repeated patterns of $\gamma$-ray flares in blazars strengthens the statistical significance of their association and supports the hypothesis that these phenomena are physically related. If these repeated patterns do indeed reflect the presence of a spine-sheath structure in jets that facilitates neutrino production, as predicted by \citet{Tavecchio2014}, several longstanding questions could be naturally resolved. Given the events discussed in this work, repeated patterns have been detected in only 11\% of $\gamma$-ray bright blazars, suggesting that only a subset of jets exhibit the spine-sheath structure necessary for efficient neutrino production. This limited occurrence could explain the inconclusive results of broader studies investigating associations between neutrinos and complete samples of radio- or $\gamma$-ray blazars \citep{Abbasi2023a}. The hypothesis linking structured jets to neutrino production also provides insight into why neutrino detections are not always accompanied by high $\gamma$-ray flux from the associated blazars. For instance, IC190730A occurred during a low $\gamma$-ray state of PKS 1502+106 (see Fig.~\ref{fig:curve_match}). The repeated patterns are likely related to the propagation of moving features in the jet through regions of the sheath that enhance the photon field \citep{Blinov2021}. In such cases, inverse Compton scattering might result only in moderate $\gamma$-ray flux, whereas brighter flares could result from different mechanisms, such as shock-shock interactions between moving and stationary jet features.

Performing detailed theoretical and numerical modeling of the repeated patterns of $\gamma$-ray flares is essential to confirm these interpretations and further improve our understanding of the underlying physics.

\begin{acknowledgements}
D.B. acknowledges support from the European Research Council (ERC) under the Horizon ERC Grants 2021 programme under the grant agreement No. 101040021.
\end{acknowledgements}

\bibliographystyle{aa}
\bibliography{bibliography.bib}

\appendix
\onecolumn
\section{Similarity of the gamma-ray light curve intervals} \label{app:A}

In this section, we provide evidence that intervals 1 and 2 of the $\gamma$-ray light curve of PKS 1502+106, shown in Fig.~\ref{fig:pattern}, are not only visually similar but also exhibit mathematical correspondence. Various measures can be used to assess similarity (distance) between light curves \citep[see e.g.][]{Lin2012}. Since the $\gamma$-ray photon flux curves analyzed here are well-sampled with regular intervals between points, we use the most common and straightforward distance measure: the Euclidean distance (ED). Assuming that light curves A and B are normalized and have the same length n, the Euclidean distance between them is defined as $ED(A,B)=\sqrt{\sum_{i=1}^n (a_i-b_i)^2}$, where $a_i$ and $b_i$ are the individual points of the corresponding curves.

\begin{figure*}[h!]
\includegraphics[width=\hsize]{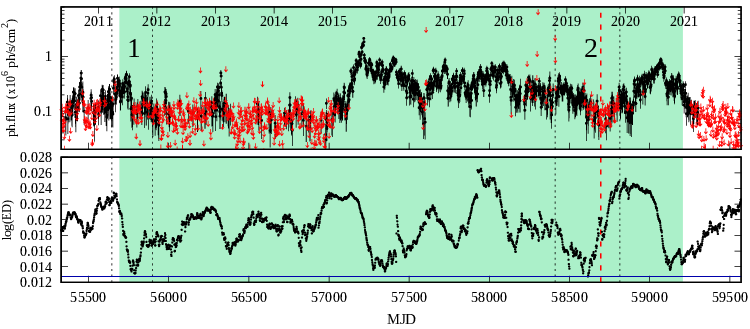}
\caption{Assessment of the randomness of similarity between two intervals of the $\gamma$-ray light curve of PKS 1502+106. Top panel: Photon flux light curve with 3.35-day binning. Bottom panel: Euclidean distance between interval 1 of the curve with a 2.11 d binning, stretched by a factor of $1/\multfact$, and the entire $\gamma$-ray light curve in the top panel after scanning across it. The numbered intervals between the vertical dashed lines represent intervals where the $\gamma$-ray curve shows the repeated pattern. The blue horizontal line corresponds to the minimal value in the Euclidean distance curve, which is located in the middle of interval 2. The red dashed line indicates the detection moment of IC190730A neutrino event.}
\label{fig:curve_match}
\end{figure*}

To demonstrate that interval 1 provides the minimal ED (i.e., is the most similar) to interval 2 along the light curve after the time scale transformation described in Sect.~\ref{sec:patt}, we performed the following exercise. For the light curve with a 2.11 d binning, we selected only the points corresponding to interval 1 (55646 $\le$ MJD $\le$ 55895). The MJDs of these selected points were then multiplied by a factor of $1/\multfact$, as outlined in Sect.~\ref{sec:patt}.  Next, we iterated along the $\gamma$-ray light curve with a 3.35 d binning, calculating the ED between it and the transformed interval 1. The result of this procedure is shown in the bottom panel of Fig.~\ref{fig:curve_match}.
The minimal ED is achieved in the middle of interval 2. Similar results were obtained using light curves with different binning, provided the ratio between the time scales of intervals 1 and 2 was kept equal to $\multfact$. Therefore, we conclude that the similarity between the analyzed intervals of the light curve is the strongest among all possible comparisons within the studied time range.

An alternative possibility is that many random intervals of the $\gamma$-ray light curve, under some transformation of the time scale, could provide relatively good similarity with interval 1. To test this, we estimated the probability that interval 2, under the identified transformations, is just accidentally close to interval 1 in Euclidean distance. For this purpose, we performed a Monte-Carlo simulation that randomly selected a portion of the light curve within the whole range shown in Fig.~\ref{fig:curve_match} and a random timescale transformation factor. The selected interval's time scale was then transformed using this factor and shifted to the Julian Date range of interval 1. The ED was calculated as before. The timescale factor was allowed to vary within the range [0.5,~2], which includes the identified factor $\multfact$. After performing $10^6$ simulations, we found that a random interval provided a better fit (i.e., a smaller ED) to interval 1 in only 50 trials, resulting in an estimated probability of random coincidence of $p=5\times10^{-5}$.
The two tests described in this section demonstrate that the similarity seen among the $\gamma$-ray light curve segments in intervals 1 and 2 is unlikely to be random.

\end{document}